 \documentclass[authoryear,preprint,review,12pt]{elsarticle}

\newcommand{\kms}{\mbox{km s$^{-1}$}}

\usepackage{graphics}

\usepackage{graphicx}

\usepackage{epsfig}

\usepackage{amssymb}

\usepackage{amsthm}

\usepackage{lineno}

\journal{New Astronomy}

\begin{document}

\begin{frontmatter}



\title{BD\,+36 3317: An Algol Type Eclipsing Binary in Delta Lyrae Cluster}


\author{O. \"{O}zdarcan\corref{cor1}}
\cortext[cor1]{Corresponding author}
\ead{orkun.ozdarcan@ege.edu.tr}

\author{E. Sipahi}

\author{H. A. Dal}

\address{Ege University, Science Faculty, Department of Astronomy and Space Sciences, 35100 Bornova, \.{I}zmir, Turkey}

\begin{abstract}
In this paper, we present standard Johnson $UBV$ photometry of the eclipsing binary 
BD\,+36 3317 which is known as a member of Delta Lyrae (Stephenson\,1) cluster. 
We determined colors and brightness of the system, calculated $E(B - V)$ color excess. 
We discovered that the system shows total eclipse in secondary minimum. Using this 
advantage, we found that the primary component of the system has $B9-A0$ spectral type. 
Although there is no published orbital solution, we tried to estimate the physical properties 
of the system from simultaneous analysis of $UBV$ light curves with 2003 version of 
Wilson - Devinney code. Then we considered photometric solution results together 
with evolutionary models and estimated the masses of the components as 
$M_{1}$ = 2.5 $M_{\odot}$ and  $M_{2}$ = 1.6 $M_{\odot}$. Those estimations 
gave the distance of the system as 353 pc. Considering the uncertainties
in distance estimation, resulting distance is in agreement with the distance 
of Delta Lyrae cluster.
\end{abstract}


\end{frontmatter}


\section{Introduction}\label{S1}

BD\,+36 3317 (SAO\,67556, $\alpha_{2000}$ = 18$^{h}$~54$^{m}$~22$^{s}$, 
$\delta_{2000}$ = 36$^{\circ}$~51$'$~07$''$, $V$ = 8$^{m}$.77) is an Algol type 
eclipsing binary which is in the same area with Delta Lyrae cluster 
($\alpha_{2000}$ = 18$^{h}$~54$^{m}$, $\delta_{2000}$ = 36$^{\circ}$~49$'$, \citet{khar05})  
in the sky. The system is considered as a member of the cluster \citep{Eg72,Eg83,AT84}. 
The existence of the cluster was suggested by \citet{Ste59} for the first time. Further photometric 
evidence for the existence of the cluster was provided by \citet{Eg68}. However, possible 
members of the cluster, included BD\,+36 3317, have considerably different distance 
modulus values \citep{AT84} and this makes harder to confirm real members, even 
the existence of cluster. After all, mean distance modulus for the cluster is given as 7$^{m}$.29 
with $E(B-V)$ = 0$^{m}$.04 by \citet{AT84}. More recent distance modulus value was given by 
\citet{khar05} as 7$^{m}$.98 (their $E(B - V)$ value is 0$^{m}$.04, too) with the distance of 373 pc. 
\citet{khar04} calculated the Delta Lyrae membership probability of BD\,+36 3317 as 
0.67, 0.91 and 1.0, in terms of proper motion, photometry and spatial position, respectively. 
This calculation seems to support the membership of BD\,+36 3317 to the cluster.

Spectroscopic binary nature of the star has been first noticed by \citet{Eg68} via its 
large radial velocity variation from $-$90 {\kms} to 17 {\kms}. However, eclipses of the 
system has been discovered by \citet{VBAH08}, many years after the discovery of its 
spectroscopic binary nature. No orbital solution via radial velocity measurements 
has been published up to know. At that point, BD\,+36 3317 has the advantage 
of to be an eclipsing binary in terms of estimating its absolute physical properties and 
distance. Furthermore, one can calculate its distance and compare it with the distance 
of the cluster and check the membership of the system to the cluster. For those purposes, 
we obtained standard Johnson $UBV$ observations of the system in 2008 and 2009. 
By using standard photometric data, we calculated $E(B-V)$ value of the system and 
estimated their spectral type. Although the lack of spectroscopic mass ratio and orbital 
solution is a disadvantage in terms of determination of the absolute dimension of the system, 
we made a fair estimation for physical properties of BD\,+36 3317, including the distance 
of the system, by using the photometric solution. In the next section, we give summary 
of our observations. In section 3 we refine the light elements of the system via $O-C$ analysis. 
In section 4, we give basic photometric properties of the system and investigate the effect 
of interstellar extinction. In section 5 we present the simultaneous analysis of $UBV$ light 
curves and our estimation for the absolute dimension of the system. In the last section we 
discuss the results.

\section{Observations}\label{S2}

We carried out Johnson $UBV$ observations of the star at Ege University Observatory 
($EUO$). We observed the star on 41 separate nights in 2008 and 2009. Our instrumental 
setup was 0.3 m Schmidth-Cassegrain telescope equipped with uncooled SSP5 photometer. 
BD\,+36 3314 was comparison star in our observations, while BD\,+36 3313 
was check star. In many observing run, when no primary or secondary minimum occurred, 
we only performed a short observing sequence as $S-C2-C1-VVVV-S-C2-C1-VVVV-C1-C2-S$, 
where $S$ is sky, $C2$ is check star, $C1$ is comparison star and $V$ is variable star. 
For those kind of nights, we used average atmospheric extinction coefficients of $EUO$ 
in order to correct all differential magnitudes in terms of $V-C1$, $V-C2$ and $C2-C1$. 
For other nights, when a primary or a secondary minimum occurred, we performed an all night 
observing run. For those nights, we calculate atmospheric extinction coefficients via 
measurements of $C1$ and made all corrections for atmospheric extinction on differential 
magnitudes according to those coefficients. We estimate average standard deviations of 
observations from $C2-C1$ measurements and resulting average standard deviations are 
0$^{m}$.053, 0$^{m}$.019 and 0$^{m}$.017 for $U$, $B$ and $V$ filters, respectively. 
We observed 11 stars from IC\,4665 cluster \citep{MM96} on 29$^{th}$ July 2008 and 9 stars 
from the list of \citet{and95} on 17$^{th}$ August 2009, in order to calculate coefficients of the 
transformation of the instrumental system to the standard one. By those coefficients, we applied 
color corrections for all differential measurements. Then, we directly calculated average 
standard magnitudes and colors of the comparison star from those two nights. Finally, by using 
standard magnitude and colors of the comparison star, we calculate standard magnitude and 
colors of the variable.

\section{$O-C$ Analysis}\label{S3}

It is not possible to make a comprehensive $O-C$ analysis for BD\,+36 3317 since 
there is not enough minimum time observations. Only the first ephemeris of the system
is available in literature \citep{VBAH08} as

\begin{equation}\label{E1}
(HJD)_{Min I} = 2,454,437.25921 + 4^{d}.30216 \ \times \ E \ .
\end{equation}

where $(HJD)_{Min I}$ is epoch, which corresponds to a time of a primary minimum of 
BD\,+36 3317, and $E$ is integer eclipse cycle number. We have already had 
three primary minima (Type I) and one secondary minimum (Type II) in our 
observations \citep{Sip09}. In Table~\ref{T1}, we list those minima with 
corresponding $O-C$ values.

\begin{table}[!htb]
\caption{The times of light minima of BD\,+36 3317. In the first column, 
the errors are given for the last digit of the measurements.}\label{T1}
\begin{center}
\begin{tabular}{|c|c|r|c|c|}
\hline
HJD         & E     &   $O-C$    & Filter    &   Type \\
(24 00000 +)  &                  &    (day)           &           &\\
\hline
54652.3522(4)  & 50.0   & $-$0.0025    & $UBV$  & I   \\
54667.4148(5)  & 53.5   &      0.0026    & $UBV$  & II  \\
55052.4561(2)  & 143.0 &      0.0004    & $UBV$  & I   \\
55078.2683(4)  & 149.0 & $-$0.0004    & $UBV$  & I   \\
\hline
\end{tabular}
\end{center}
\end{table}

Application of linear least squares method to primary and secondary minima data gives 
very small amount of corrections in the ephemeris. The new light elements and 
their errors are as follows:

\begin{equation}\label{E2}
(HJD)_{Min I} = 2,454,437.2466(30) + 4^{d}.302162(27) \ \times \ E \ .
\end{equation}

For further analysis of light curves, we use those final light elements.

\section{Basic Photometric Parameters and Interstellar Extinction}\label{S4}

\begin{figure*}[ht!]
\begin{center}\includegraphics[angle=0,height=8.8cm,width=12cm]{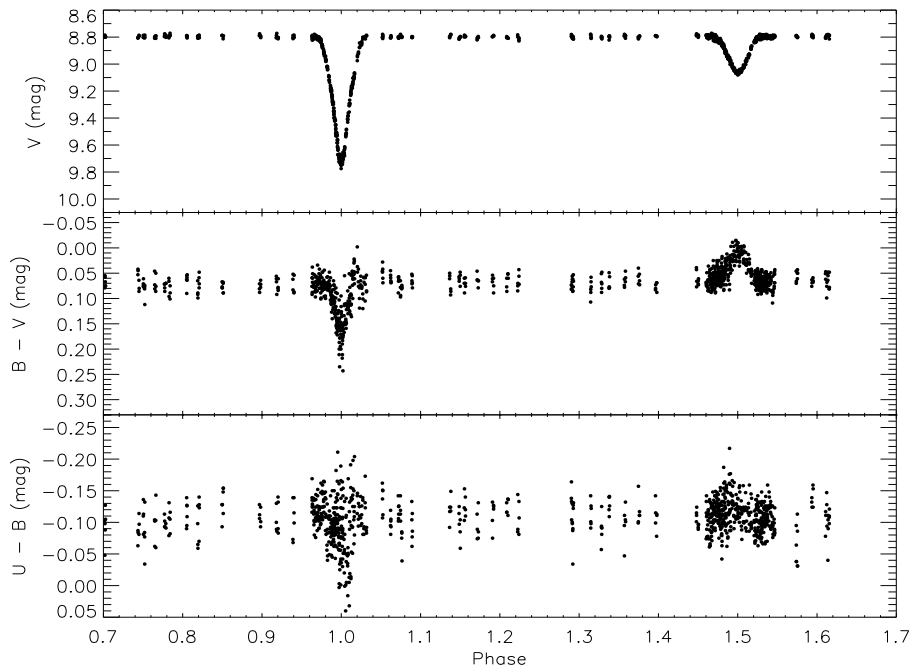}\end{center}
\caption{Phase folded and standardized light and color curves of BD\,+36 3317.} \label{F1}
\end{figure*}

We show phased light and color curves of the system in Figure~\ref{F1}. 

One can easily notice clear variations in $B - V$ in primary and secondary minima, 
which indicate quite large temperature difference among the components of the 
system. We determined brightnesses and colors of the system in maximum light, 
primary and secondary minima, which would give us some hints about components. 
We list those values in Table~\ref{T2}.

\begin{table}[ht!]
\caption{Magnitude and colors of BD\,+36 3317.}\label{T2}
\begin{center}
\begin{tabular}{|c|c|r|c|c|}
\hline
      &  Max   & Min I   & Min II  \\
      & (mag)  & (mag)  & (mag)  \\
\hline
V	   &    8.798 &	9.715 &  9.053 \\
U-B &   -0.108 &  -0.090 & -0.119 \\
B-V &    0.066 &	0.167 &  0.009 \\
\hline
\end{tabular}
\end{center}
\end{table}

We can inspect effect of interstellar extinction via $UBV$ color - color diagram, 
by considering the colors at maximum light. We used $UBV$ standard star data 
from \cite{DL00} for this purpose. First, we used magnitude and colors of maximum 
light (see $Max$ column in Table~\ref{T2}) by assuming $E(U-B)/E(B-V) = 0.72$ as 
the slope of the reddening vector in $UBV$ color - color diagram. Resulting 
$E(B - V)$ and $A_{v}$ values are 0$^{m}$.139 and 0$^{m}$.43, respectively, under the assumption 
of $R$ = 3.1. Then, de-reddened total color for the system is $(B - V)_{0}$ = -0$^{m}$.07. 
Now we can make an estimation for the temperature of the primary component, 
by assuming that the maximum contribution to the total light comes from the 
primary component. However, during the light curve analysis, we noticed that
orbital inclination value of the system shows very small variations around 89.4 degrees which 
makes the secondary minimum total eclipse. Then, the secondary component is completely 
hidden behind the primary component at the middle of the secondary minimum, hence, 
the magnitude and colors of that phase corresponds to the direct measurements of 
the primary component. Here, we refer the reader to the next section for justification of 
$''total~eclipse~at~secondary~minimum''$ case. In this case, we have direct measurements 
of the primary component (third column in Table~\ref{T2}) and total colors and magnitude of the system 
(first column in Table~\ref{T2}), therefore we can easily calculate magnitude and colors of the 
secondary component as $V$ = 10$^{m}$.497, $U-B$ = $-$0$^{m}$.053 and $B-V$ = 0$^{m}$.316. 
This case is also an advantage in terms of more accurate determination of the interstellar 
reddening and intrinsic colors of the components separately. Therefore, we repeated the 
calculation of the interstellar reddening as described above, but this time by using the direct 
measurements of the primary component at the middle of the secondary minimum. 
We list intrinsic colors of the components together with the more accurate interstellar 
reddening and extinction estimation in Table~\ref{T3}.

\begin{table}[ht!]
\caption{Intrinsic colors and magnitude of the components of BD\,+36 3317 
together with the amount of interstellar reddening and extinction.}\label{T3}
\begin{center}
\begin{tabular}{|c|c|c|}
\hline
      &  Primary  & Secondary                    \\
      &   (mag)    &    (mag)                    \\
\hline
$V_{0}$	   &    8.833 &	10.277             \\
$(U-B)_{0}$ &   -0.170 &  -0.104             \\
$(B-V)_{0}$ &   -0.062 &	0.245             \\
\hline
$E(B-V)$     &   \multicolumn{2}{|c|}{0.07}                  \\
$A_{v}$      &  \multicolumn{2}{|c|} {0.22}                   \\
\hline
\end{tabular}
\end{center}
\end{table}

According to Table~\ref{T3}, colors of the primary component indicates $B9$ spectral 
type while secondary components corresponds about $A8$ spectral type.


\section{Analysis of Light Curves}\label{S5}

Under the assumption of total eclipse in secondary minimum, we can estimate the effective 
temperature of the primary component directly, which is very critical for light curve 
analysis. We adopted calibration of \citet{G05} for effective temperature estimation 
of the primary component and resulting temperature for $(B - V)_{0}$ = $-$0$^{m}$.062 
is $T_{1}$ = 10750 $K$ with the error of $\sigma_{T_{1}}$ = 470 $K$. The error of 
the temperature is calculated from the standard deviation of $B - V$ color in our observations. 

\begin{figure}[!ht]
\includegraphics[angle=0,width=\textwidth]{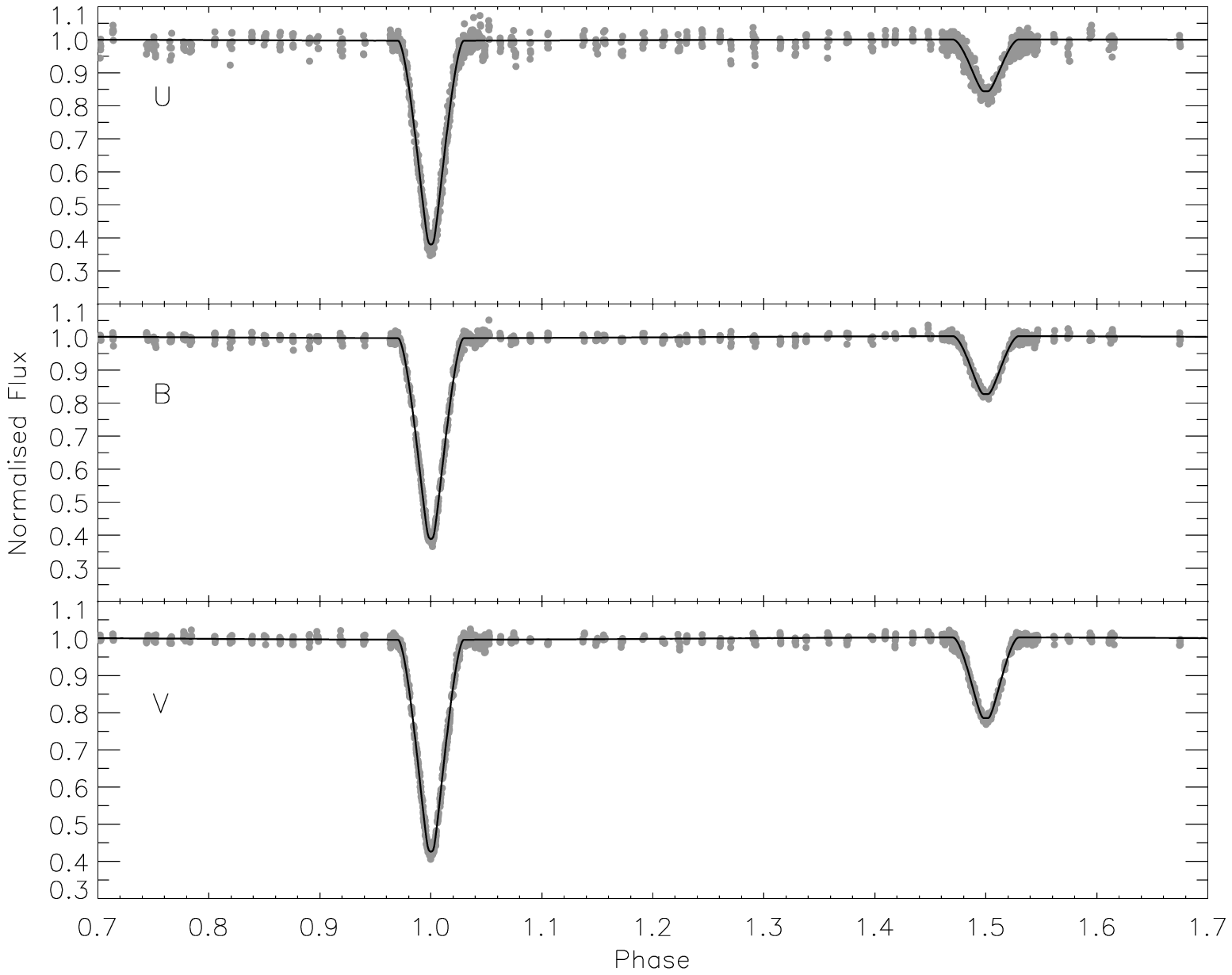}
\caption{Representation of observational data (points) in terms of normalised 
flux and theoretical solution (continuous line).}\label{F2}
\end{figure}

Before starting analysis, we chose 0.25 phase as normalisation phase and converted 
all magnitude measurements into normalised flux according to the light level at that phase. 
For light curve analysis, we used 2003 version of the Wilson - Devinney code 
\citep{WD71,W79,W90}. Photometric properties of the components gives us hints for 
some parameters to reduce the number of free parameters in photometric solution. 
We adopt the gravity brightening coefficients $g_{1}$ = $g_{2}$ = 1 and albedos 
$A_{1}$ = $A_{2}$ = 1 for stars which have radiative envelopes. We assume synchronised 
rotation for both components, so $F_{1}$ = $F_{2}$ = 1. We took the band-pass dependent 
($x_{1,2}$, $y_{1,2}$) and bolometric ($x_{1,2}(bol)$, $y_{1,2}(bol)$) limb darkening coefficients 
from \citet{VH93} by assuming square root law \citep{DCG92} which is more appropriate 
for stars hotter than 8500 $K$. Since there is no radial velocity study in literature, we do 
not have any information about mass ratio, which is another critical parameter for light 
curve analysis. Although it is not an efficient way to determine $q$ from photometry in 
detached systems, we searched for the best solution for different $q$ values, starting from 
$q$ = 0.30 until $q$ = 0.90, by using $UBV$ data simultaneously. Orbital inclination ($i$), 
effective temperature of the secondary component ($T_{2}$), $\Omega$ potentials of 
primary and secondary ($\Omega_{1}$, $\Omega_{2}$) and luminosity of the primary 
component ($L_{1}$) are free parameters in the solution. The errors of the best solutions 
for individual $q$ values are very close to each other between $q$ = 0.45 and $q$ = 0.70. 
At that point, we considered mass - luminosity relation as $L\propto M^{4}$ by using the resulting
absolute luminosities at the end of the solution (see later results in this section). 
We repeated solutions for many $q$ values between $q$ = 0.45 and $q$ = 0.70. 
In most cases, $L\propto M^{4}$ relation indicates the $q$ value close to 0.65, hence, 
we finally accepted $q$ = 0.65 and applied a final light curve solution. In Table~\ref{T4}, 
we give final light curve analysis results. We note that the error of $T_{2}$ is internal to the 
Wilson - Devinney code and its error should be similar to the error of $T_{2}$. 
In the table, $\langle r_{1,2}\rangle$ denotes average of three fractional radius values 
($pole$, $side$ and $back$ values in solution output) relative to the semi-major axis of the orbit, 
for corresponding component. We give the normalised fluxes for $U$, $B$, and $V$ filters and 
corresponding theoretical light curves for our final solution in Figure~\ref{F2}.

\begin{table}[!ht]
\caption{Light curve analysis results of BD\,+36 3317. Errors are given in parenthesis.}\label{T4}
\begin{center}
\begin{tabular}{|c|c|}
\hline
$q( = M{_2}/M{_1})$ &  0.65 (fixed)\\
$T_{1}(K)$ &  10750 (fixed)\\
$g_{1}$ = $g_{2}$  & 1.0\\
$A_{1}$ = $A_{2}$  & 1.0\\
$F_{1}$ = $F_{2}$  & 1.0\\
$i~(^{\circ})$ &  89.61(11)\\
$T_{2}(K)$ &  7711 (10)\\
$\Omega_{1}$ & 10.571 (25)\\
$\Omega_{2}$ &   8.997 (23)\\
$L_{1}$/($L_{1}$+$L_{2})_{U}$ & 0.851(17) \\
$L_{1}$/($L_{1}$+$L_{2})_{B}$ & 0.831(15) \\
$L_{1}$/($L_{1}$+$L_{2})_{V}$ & 0.789(15) \\
$x_{1}(bol),x_{2}(bol)$ &  0.558, 0.215\\
$y_{1}(bol),y_{2}(bol)$ &  0.172, 0.525\\
$x{_1}, y{_1} (U)$ &  0.082 , 0.590 \\
$x{_1}, y{_1} (B)$ & $-$0.058 , 0.846 \\
$x{_1}, y{_1} (V)$ & $-$0.047 , 0.723 \\
$x{_2}, y{_2} (U)$ &  0.184 , 0.646 \\
$x{_2}, y{_2} (B)$ &  0.105 , 0.822 \\
$x{_2}, y{_2} (V)$ &  0.096 , 0.711 \\
$\langle r_{1}\rangle, \langle r_{2}\rangle$ & 0.1009(3), 0.0832(2)\\
rms & 0.011 \\
\hline
\end{tabular}
\end{center}
\end{table}

\begin{figure}[!ht]
\includegraphics[angle=0,width=\textwidth]{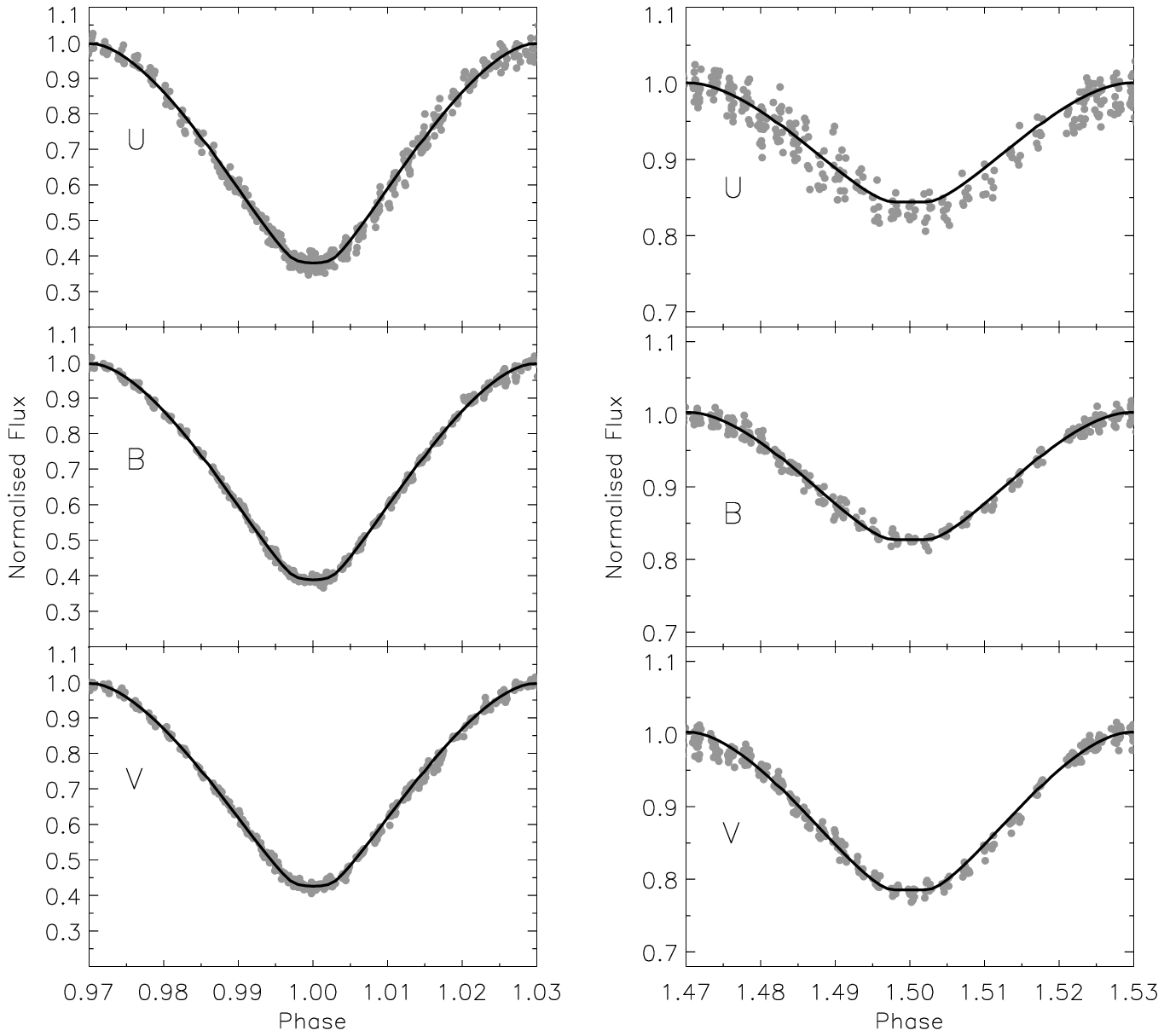}
\caption{Close look to the primary and secondary minima in each filter. In secondary 
minimum, one can notice the short total eclipse. }\label{F3}
\end{figure}

In Figure~\ref{F3}, we zoom to the primary (left panels) and secondary (right panels) 
minima to show the shapes of the eclipses. One can notice that the secondary 
minimum is certainly total eclipse which lasts for a short phase range. At the primary
minimum, we observe non-flat bottomed light variation which shows the
effects of annular eclipse and limb darkening together.

Comparison among photometric solution results, mass - luminosity relation and 
evolutionary models of \citet{Gir00} indicates that the mass of the primary $M_{1}$ 
is close to the 2.5 $M_{\odot}$, therefore we assume $M_{1}$ = 2.5 $M_{\odot}$. 
This assumption makes the mass of the secondary component $M_{2}$ = 1.6 $M_{\odot}$ 
according to the $q$ value.

Since we estimated the masses of the components, we can go one step further 
and calculate the absolute parameters and the distance of the system. By the 
aid of Kepler's third law, we can calculate semi-major axis of the orbit $a$. After that 
point, one can easily calculate absolute radii of the components by using average 
fractional radii in Table~\ref{T4}. Now we have effective temperatures and absolute 
radii of the components and we can calculate their luminosities in solar unit via 
$Stefan - Boltzmann$ law, by using $T_{\odot}$ = 5770 K. We list absolute 
parameters of the system in Table~\ref{T5}. Finally, we can calculate the distance 
of the system by using primary and secondary component separately, via their 
photometric and absolute properties. In distance calculation, we adopted bolometric 
corrections from \cite{G05}. Photometric properties of the primary component leads 
to a distance of 353 pc. We only use primary component in order to estimate the distance
since its signal is very strong relative to the secondary component.

\begin{table}[!ht]
\caption{Absolute Physical Properties of BD\,+36 3317. Solar $M_{bol}$ 
value of 4.$^{m}$74 is used to calculate $M_{bol}$ values of the components.}\label{T5}
\begin{center}
\begin{tabular}{|c|c|c|}
\hline
Parameter  & Primary  & Secondary  \\
\hline
Mass $(M_{\odot})$      &  2.5  & 1.6\\
Radius $(R_{\odot})$     &  1.8    &1.5 \\
Luminosity $(L_{\odot})$ &  39   & 7 \\
$M_{bol}$ (mag)    &  0.76    & 2.62 \\
$log(g)$ (cgs)     &  4.32    &4.31 \\
\hline
\end{tabular}
\end{center}
\end{table}

\section{Summary and conclusions}\label{S6}

We presented standard Johnson $UBV$ photometry of the Algol type
eclipsing binary BD\,+36 3317 with a fairly good phase coverage and 
reasonably accurate observational data. During observations, we obtained
three primary and one secondary minima. We refined the epoch and the 
period of the system by applying linear least squares method to the timings of
those light minima. We determined standard colors and magnitude of the system in
maximum light, primary minimum and secondary minimum. During the analysis,
we noticed that the secondary minimum is total eclipse, which is an advantage in analysis 
and means that the measurement at that phase corresponds to direct measurements 
of the primary component. Using this advantage, we first determined interstellar
reddening and extinction via direct measurements of the primary component. 
Then, we were able to calculate de-reddened colors and magnitudes of the 
components, separately. This case was an another advantage in order to make a 
more accurate estimation of $T_{1}$, hence enabled us to reduce the number of 
free parameters more reliably in simultaneous $UBV$ light curve solution.
Photometric solution justified that the secondary minimum was total eclipse.

Lack of an orbital solution based on radial velocity measurements prevented us
from determining the accurate absolute parameters of the system and their uncertainties. 
Hence, we can only make a rough estimation for the uncertainties and check how those
uncertainties affect our results.

If we assume that the $q$ is between 0.6 and 0.7 and $M_{1}$ is between
2.4 $M_{\odot}$ and 2.6 $M_{\odot}$, then we can calculate a lower and 
upper limit for $a$ via Kepler's third law. Those lower and upper limits of $a$ 
put $R_{1}$ between 1.76 $R_{\odot}$ - 1.84 $R_{\odot}$ when we consider the
average fractional radius of the primary component from photometric solution. 
Same calculation puts $R_{2}$ between 1.45 $R_{\odot}$ - 1.52 $R_{\odot}$. A similar 
method can be used for the luminosities of the components by using 
$Stefan - Boltzmann$ law and $T_{\odot}$ = 5770 K. Assuming a lower and 
upper limits for $T_{1}$ and  $T_{2}$ via $\sigma_{T_{1}}$ = 470 $K$, 
we can calculate the ranges of $L_{1}$ and $L_{2}$ as 
31 $L_{\odot}$ - 49 $L_{\odot}$ and  5 $L_{\odot}$ - 9 $L_{\odot}$, respectively.

\begin{figure}[!ht]
\begin{center}
\includegraphics[angle=0,height=9cm,width=9cm]{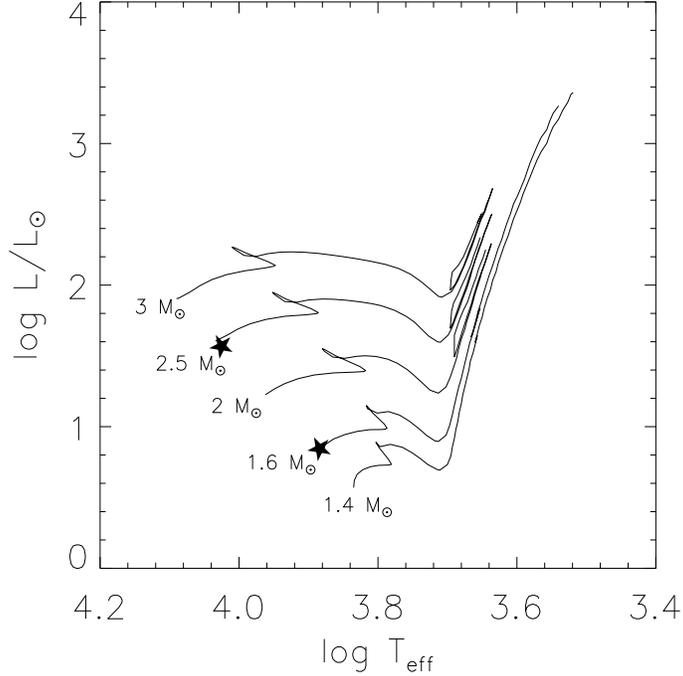}\end{center}
\caption{Positions of the components (filled stars) in log $T_{eff}$ - log $L$ plane. 
All tracks for $Z$ = 0.019 are from \citet{Gir00}.}\label{F4}
\end{figure}

We show preliminary plots of the components in log $T_{eff}$ - log $L$ plane 
in Figure~\ref{F4}. Evolutionary tracks for solar abundance ($Z$ = 0.019) comes from 
\citet{Gir00}. The components of the system seem close to the ZAMS and in good 
agreement with solar metal abundance. However, spectroscopic analysis is necessary 
to revise or refine it.

We can make an estimation for the uncertainty of the distance in a similar way,
as described above. If we assume the same ranges for $q$, $M_{1}$ and $T_{1}$ as 
in previous uncertainty estimations, we can calculate the range of absolute bolometric 
magnitude of the primary component as 1$^{m}$.01 - 0$^{m}$.52. Here, we adopt 
bolometric corrections from \cite{G05} for corresponding temperature limits in order to 
calculate limit visual absolute magnitudes ($M_{V}$) of the primary component. 
Using de-reddened $V$ magnitude and $M_{V}$ limits of the primary component in 
distance modulus, we can calculate the range of the distance as 329 - 377 pc and 
this leads a mean value of 353 pc which is our estimation in previous section. 
The uncertainty in the distance might be slightly exaggerated since we make a rough 
uncertainty estimation for related parameters. However, our distance estimation differs 
only by 20 pc from distance value of the cluster \citep{khar05}. This confirms the 
membership of BD\,+36 3317 to Delta Lyrae cluster. Nevertheless, comprehensive 
spectroscopic study of the system, in terms of radial velocity measurements, would 
help to refine or revise the physical properties and distance of the system. Further 
spectroscopy would also give a chance to check the metal abundance of the system 
which also contains some hints about the nature of the cluster.



\section*{Acknowledgments} The authors acknowledge generous allotments of observing time at the Ege University Observatory.


\begin{thebibliography}{16}
\bibitem[Andruk et al.(1995)]{and95}Andruk, V., Kharchenko, N., Schilbach, E., Scholz, D., 1995, AN, 316, 225
\bibitem[Anthony-Twarog(1984)]{AT84}Anthony-Twarog, B.J., 1984, AJ, 89, 655
\bibitem[D\'{i}az-Cordov\'{e}s and Gim\'{e}nez(1992)]{DCG92}D\'{i}az-Cordov\'{e}s J., Gim\'{e}nez A., 1992, A\&A, 259, 227
\bibitem[Drilling and Landolt(2000)]{DL00}Drilling, J. S., Landolt, A. U., 2000, in Cox A. N. (ed.),\emph{Allens Astrophysical Quantities, 4th ed.}, Springer, Berlin, p.~388
\bibitem[Eggen(1968)]{Eg68}Eggen, O.J., 1968, ApJ, 152, 77
\bibitem[Eggen(1972)]{Eg72}Eggen, O.J., 1972, ApJ, 173, 63
\bibitem[Eggen(1983)]{Eg83}Eggen, O.J., 1983, MNRAS, 204, 391
\bibitem[Girardi et al.(2000)]{Gir00}Girardi, L., Bressan, A., Bertelli, G., Chiosi, C., 2000, A\&AS, 141, 371
\bibitem[Gray(2005)]{G05}Gray, D.F., 2005, \emph{The observation and analysis of stellar photospheres}, 3rd ed., Cambridge Univ. Press
\bibitem[Kharchenko et al.(2004)]{khar04}Kharchenko, N.V., Piskunov, A.E., R\"oser, S., Schilbach, E., Scholz, R.-D., 2004, AN, 325, 740
\bibitem[Kharchenko et al.(2005)]{khar05}Kharchenko, N.V., Piskunov, A.E., R\"oser, S., Schilbach, E., Scholz, R.-D., 2005, A\&A, 438, 1163
\bibitem[Menzies and Marang(1996)]{MM96}Menzies, J. W., Marang, F, 1996, MNRAS, 282, 313
\bibitem[Sipahi et al.(2009)]{Sip09}Sipahi, E., Dal, H.A., \"Ozdarcan, O., 2009, IBVS, 5904, 1
\bibitem[Stephenson(1959)]{Ste59}Stephenson, C.B., 1959, PASP, 71, 145
\bibitem[van Hamme(1993)]{VH93}van Hamme, W., 1993, AJ, 106, 2096
\bibitem[Violat-Bordonau and Arranz Heras(2008)]{VBAH08}Violat-Bordonau, F., Arranz-Heras, T., 2008, IBVS, 5900, 7
\bibitem[Wilson and Devinney(1971)]{WD71}Wilson, R.E., Devinney, Edward J., 1971, ApJ, 166, 605
\bibitem[Wilson(1979)]{W79}Wilson, R.E., 1979, ApJ, 234, 1054
\bibitem[Wilson(1990)]{W90}Wilson, R.E., 1990, ApJ, 356, 613
\end{thebibliography}
\end{document}